\RequirePackage{cmap}
\documentclass[aps,pra,onecolumn,superscriptaddress]{revtex4-1}

\usepackage[T1]{fontenc}
\usepackage[utf8]{inputenc}
\usepackage{graphicx}
\usepackage{amsmath}
\usepackage{amsfonts}
\usepackage{amssymb}
\usepackage{epsfig}
\usepackage{color}
\usepackage{setspace}
\setcitestyle{super}
\makeatletter
\newcommand*{\citenumns}[2][]{%
  \begingroup
  \let\NAT@mbox=\mbox
  \let\@cite\NAT@citenum
  \let\NAT@space\NAT@spacechar
  \let\NAT@super@kern\relax
  \renewcommand\NAT@open{}%
  \renewcommand\NAT@close{}%
  \cite[#1]{#2}%
  \endgroup
}
\makeatother

\newcommand\eal{\langle\!\langle}
\newcommand\ear{\rangle\!\rangle}
\def\<{\langle}
\def\>{\rangle}
\newcommand\opone{\leavevmode\hbox{\small1\kern-3.8pt\normalsize1}}
\newcommand{\e}{{\rm e}}
\DeclareMathOperator{\Tr}{Tr}
\AtBeginDocument{}
\usepackage{url}
\expandafter\def\expandafter\UrlBreaks\expandafter{\UrlBreaks%  save the current one
  \do\a\do\b\do\c\do\d\do\e\do\f\do\g\do\h\do\i\do\j%
  \do\k\do\l\do\m\do\n\do\o\do\p\do\q\do\r\do\s\do\t%
  \do\u\do\v\do\w\do\x\do\y\do\z\do\A\do\B\do\C\do\D%
  \do\E\do\F\do\G\do\H\do\I\do\J\do\K\do\L\do\M\do\N%
  \do\O\do\P\do\Q\do\R\do\S\do\T\do\U\do\V\do\W\do\X%
  \do\Y\do\Z\do\1\do\2\do\3\do\4\do\5\do\6\do\7\do\8%
  \do\9\do\0}
	
\begin{document}

\title{Eternal non-Markovianity: from random unitary to Markov chain realisations}

\author{Nina Megier}
\email{nina.megier@tu-dresden.de}
\affiliation{Institut f\"ur Theoretische Physik, Technische Universit\"at Dresden, D-01062 Dresden, Germany}
\author{Dariusz Chru\'sci\'nski}
\affiliation{Institute of Physics, Faculty of Physics, Astronomy and Informatics, Nicolaus Copernicus University, Grudzi\k{a}dzka 5/7, 87-100 Toru\'n, Poland}
\author{Jyrki Piilo}
\affiliation{Turku Centre for Quantum Physics, Department of Physics and Astronomy, University of Turku, FI-20014 Turun Yliopisto, Finland }
\author{Walter T. Strunz}
\affiliation{Institut f\"ur Theoretische Physik, Technische Universit\"at Dresden, D-01062 Dresden, Germany}

\date{\today}
\maketitle

\section*{Abstract}
The theoretical description of quantum dynamics in an intriguing way does not necessarily imply
the underlying dynamics is indeed intriguing. Here we show how a known very interesting master
equation with an always negative decay rate [{\it eternal non-Markovianity} (ENM)]
arises from simple stochastic Schr\"odinger dynamics (random unitary dynamics).
Equivalently, it may be seen as arising from a mixture of Markov (semi-group) open system dynamics.
Both these approaches lead to a more general family of CPT maps, characterized by a point
within a parameter triangle. Our results show how ENM quantum dynamics can be realised easily in the laboratory. 
Moreover, we find a quantum time-continuously measured (quantum trajectory) realisation
of the dynamics of the ENM master equation based on unitary transformations and projective measurements in an extended Hilbert space,
guided by a classical Markov process. Furthermore, a Gorini-Kossakowski-Sudarshan-Lindblad (GKSL) representation of the dynamics in an extended Hilbert space
can be found, with a remarkable property: there is no dynamics in the ancilla state. Finally, analogous constructions for two qubits  extend these results 
from non-CP-divisible to non-P-divisible dynamics.

\section*{Introduction}
A realistic modelling of many quantum phenomena inevitably needs to take into account the interaction of
our system of interest with environmental degrees of freedom. Thus, in order to describe the quantum system
dynamics appropriately, one is often forced to deal with {\it open quantum systems}.
A very relevant and well understood class of such open quantum system dynamics follows from a Markov master
equation of GKSL form~\cite{lindblad,gorini}.
Non-Markovian behaviour may arise from a structured environment or strong system-environment interaction \cite{UWeiss}.
Non-Markovian systems are very challenging: in the often-employed projection operator formalism their dynamics involves memory kernels \cite{nakajima,zwanzig}.
Other approaches range from path integrals \cite{makri,haenggi}, over hierarchical equations of motion
(HEOM) for the reduced density matrix \cite{tanimura,kreisbeck}, to hierarchies of stochastic pure states (HOPS) \cite{walter1,walter2}. Sometimes
time-convolutionless master equations can be used \cite{breuer2}. During the last few years,
due to tremendous experimental progress
in quantum technologies
in many different areas and more and more refined measurement schemes,
specific investigations of non-Markovian quantum dynamics, where GKSL is no longer
applicable, have become possible \cite{xu,LiuBH2011,Cialdi2011,Madsen,Tang2012,Groblacher}.
Recent experiments also demonstrate how to use non-Markovianity for entanglement
preservation \cite{XiangGY2014} and
for a quantum information protocol \cite{LiuBH2016}.

The theory of non-Markovian quantum dynamics is much less developed than the GKSL class and subject of
tremendous research over the last decade and more. A very valid point of view would be to
call any dynamics other than GKSL semigroup evolution ``non-Markovian''. A more detailed analysis,
however, reveals an astonishing variety of possible definitions of what constitutes non-Markovian
dynamics \cite{breuer4,rivas2,breuer3}, and therefore
a large number of definitions and measures of {\it non-Markovianity} have been proposed
\cite{wolf2,breuer,Rivas,Lu,Luo,Zhong,Liu,Lorenzo2,Bylicka2014}. So far, most studies are based on the effective dynamics of the reduced density operator,
 other consider the full dynamics of system and environment \cite{ansgar,anna}.

As mentioned earlier, in some cases of interest, the open system dynamics may be written in terms of a time-local
master equation involving time-dependent functions
as prefactors with otherwise GKSL form. Then, for some periods of time
negative decay rates may show up, which according to some measures indicates non-Markovian dynamics \cite{hallcresserandersson,laine2012jphysb}.
Recently, a remarkable master equation for a qubit was presented involving an always negative decay
rate of an otherwise GKSL-type-looking master equation. It was termed the master equation of eternal
non-Markovianity (ENM master equation) \cite{hallcresserandersson,Cresser2010}.

 We expect non-Markovian dynamics to be related to some form of memory-dependence arising from the dynamics of the environmental degrees of freedom.
 This is why non-Markovianity is associated to a "backflow of information" \cite{breuer,breuer5,jyrki,jyrki1} or to the occurrence of quantum memory \cite{breuer3}, or 
 simply to a joint
 complex system-environment dynamics \cite{modi}. In such cases, the measures detect non-Markovianity. In this contribution we want to emphasize, however, that the reverse 
 need not be true: there are non-Markovian
 master equations (according to one of the definitions), whose physical realisation does not support any notion of such "memory effects". Instead, either there is no dynamical 
 environment
 at all, the dynamics can be realised by a classical Markov process or, when embedded in a larger Hilbert space, there is no dynamics of the
 environmental state.

In this paper we derive the ENM master equation from an appropriate mixture of Markov dynamics in
two (related) ways: one is based on random unitary evolution, the second approach uses a mixture of
Markov GKSL maps. By highlighting the equivalence of all these dynamics on the reduced level,
we show explicitly how ENM evolution of a qubit could be realised in a laboratory either with a white noise or with a classical jump process with time independent jump
probabilities. Moreover, also the bipartite GKSL representation, for which the ancilla state is frozen, is possible.
Nonetheless, we may choose to describe the dynamics in terms of a negative-rate time-local master equation, or, involving a non-trivial memory integral.
These findings support the point of view that the interpretation of non-Markovianity is elusive and great care has to be taken when talking about memory effects based solely
on a reduced (master equation) description.

\section*{Time-local master equations and negative decay rates}
% \label{MARKOV VS. NON-MARKOV}

For any total Hamiltonian of system and environment and for any product initial state,
 the dynamics of an open quantum system can be expressed in terms of the dynamical map
 $\rho(t)= \Lambda_t[\rho(0)],$ with $\Lambda_t$ completely positive and trace preserving (CPT).
If $\Lambda_t$ is an invertible map then one finds the corresponding time-local  generator $\mathcal{L}_t = \dot{\Lambda}_t \Lambda_t^{-1}$  such that a time-local master
equation $\dot{\rho}(t)= \mathcal{L}_t[\rho(t)]$ follows. Assuming the semi-group property
 $\Lambda_{t+s}=\Lambda_t \Lambda_s$, the generator takes the GKSL form \cite{lindblad,gorini} ($\hbar =1$):
\begin{equation}
\dot{\rho}(t)=-i[H,\rho(t)]+\sum\limits_i \left(L_i\rho(t) L_i^+-\frac{1}{2}\{L^+_iL_i,\rho(t)\}\right),
\end{equation}
here written in a canonical form, where the $L_i$ are traceless orthonormal operators.
By any definition, dynamics described by the semigroup master equation is Markovian.

Generalised Markovian dynamics appears when the master equation takes the
quasi-GKSL-form \cite{hallcresserandersson,rivas4}
\begin{align}\label{me-gamma}
\dot{\rho}(t)&=-i[H(t),\rho(t)] \nonumber \\&+\sum\limits_i \gamma_i(t) \left(L_i(t)\rho(t) L_i^+(t)-\frac{1}{2}\{L^+_i(t)L_i(t),\rho(t)\}\right),
\end{align}
with decay rates $\gamma_i(t)\geq0$, for all $i$. Equation (\ref{me-gamma}) defines a reasonable
dynamics if applied to any state at any time and therefore defines a CP-divisible dynamical
map $\Lambda_t$ \cite{wolf}, i.e. the dynamical map $\Lambda_t$ satisfies the following property $\Lambda_t = \Lambda_{t,s} \Lambda_s$ and the family of maps (propagators) $\Lambda_{t,s}$ is CPT for any $t>s$.
It seems natural to regard dynamical maps $\Lambda_t$ with master equations of type (\ref{me-gamma})
for which $\gamma_i(t) <0$ for some $i$
and some $t$ as candidates for non-Markovian quantum dynamics. In these cases, the dynamical map is
no longer CP-divisible. Indeed, some authors \cite{hallcresserandersson} propose to use the negativity of
decoherence rates as a definition of non-Markovianity of the dynamics. This approach is
based on the fact that the canonical form of the master equation, defined in analogy to the Markov case
(so the time dependent Lindblad operators are traceless, normalized and
mutually orthogonal), is unique. Consequently, to all CPT maps generated by a master equation
of form (\ref{me-gamma}) one can uniquely assign a set of $\gamma_i(t)$.

Actually, one also considers $\Lambda_{t,s}$ which is not necessarily CP. If $\Lambda_{t,s}$ is positive for all $t>s$ then one calls the evolution P-divisible. Recently,
this notion was refined in ref. \citenumns{darek} as follows: the evolution is $k$-divisible if $\Lambda_{t,s}$ is $k$-positive. CP-divisibility is fully characterised by the
corresponding time-local generator $\mathcal{L}_t$ -- all local decoherence rates $\gamma_i(t)$ are always non-negative. P-divisibility is more difficult to characterise on
the level of the generator. One has the following property: if $\Lambda_t$ is P-divisible, then
\begin{equation}\label{P}
  \frac{d}{dt} ||\Lambda_t[X]||_1 \leq 0 ,
\end{equation}
for all Hermitian operators $X$, where $||\cdot||_1$ is a trace norm. Actually, when $\Lambda_t$ is invertible then (\ref{P}) implies P-divisibility. This property is very
close to the so-called BLP condition \cite{breuer} which says that $\Lambda_t$ defines Markovian evolution if
\begin{equation}\label{BLP}
  \frac{d}{dt} ||\Lambda_t[\rho_1-\rho_2]||_1 \leq 0 ,
\end{equation}
for all initial states $\rho_1$ and $\rho_2$. It is clear that CP-divisibility implies P-divisibility and this implies the BLP condition of information loss (\ref{BLP}).

The very insightful example of \cite{andersson,hallcresserandersson}, used throughout this work,
is the unital dynamics (i.e.: $\Lambda_t[\opone]=\opone$) of a single qubit
determined from the master equation
\begin{equation}
\dot{\rho}(t)=\frac{1}{2}\sum\limits_{k=1}^{3} \gamma_k(t)(\sigma_k\rho(t)\sigma_k - \rho(t)) \label{ru-meq},
\end{equation}
where $\sigma_k$ are the Pauli spin operators.

Defining $\lambda_i(t) =  e^{-\Gamma_j(t) - \Gamma_k(t)}$, where $\Gamma_k(t) = \int_0^t \gamma_k(u)du$, and $\{i,j,k\}$ run over the cyclic permutations of $\{1,2,3\}$, 
one has the following conditions which guarantee that the evolution $\Lambda_t$ is CPT:
\begin{equation}\label{lll}
  \lambda_i(t) + \lambda_j(t) \leq 1 + \lambda_k(t).
\end{equation}
Clearly, the corresponding dynamical map is CP-divisible iff $\gamma_k(t) \geq 0$. Interestingly, the  dynamical map is P-divisible iff the weaker conditions are satisfied 
\cite{chruscinski,chruscinski2}
\begin{equation}\label{gg}
  \gamma_i(t) + \gamma_j(t) \geq 0 \ , \ \ i\neq j ,
\end{equation}
given the validity of (\ref{lll}). Actually, in this case P-divisibility is equivalent to the BLP condition (\ref{BLP}).\\
Using the geometric measure of non-Markovianity based on the volume of admissible states \cite{Lorenzo2}, our one qubit dynamics is classified as Markov, too, as for all
times $\gamma_1(t) + \gamma_2(t) +\gamma_3(t)>0$ \cite{chruscinski} is satisfied.

An interesting example of the generator was proposed in ref. \citenumns{hallcresserandersson} - the ENM master equation, with
\begin{align}
\gamma_1(t)=\gamma_2(t)=1, && \gamma_3(t)=-\tanh(t), \label{gammas}
\end{align}
where one rate is always negative: $\gamma_3(t) < 0$ for all $t>0$.
One easily checks that (\ref{lll}) are satisfied and hence the dynamical map is CPT. Clearly, the corresponding dynamical map  is not CP-divisible because
of the negativity of $\gamma_3(t)$. Moreover, conditions (\ref{gg}) are also satisfied which implies that the map is P-divisible \cite{chruscinski,chruscinski2}.

Is this evolution non-Markovian? Based on the concept of CP-divisibility it is clearly non-Markovian. However,
it satisfies condition (\ref{BLP}), hence it is Markovian according to BLP.
In the following we want to argue that the meaning of non-Markovianity for non-CP-divisible maps like those
generated by (\ref{me-gamma}) with an always negative rate (\ref{gammas}) needs to be discussed carefully. 
In particular, it can be highly misleading here to relate the formal property of ``non-Markovianity'' according to one of its definitions
to some notion of ``complex system-environment dynamics'' or ``backflow of information'' from environment to system as
will be exemplified in this paper.

We show that there is a whole family of master equations of type (\ref{me-gamma}) with $\gamma_i(t) <0$,
for some $i$ and times $t$ that
i) turn out to arise from random unitary Schr\"odinger dynamics, ii) are mere mixtures of Markovian semi-group dynamics, iii) allow for a physical realisation based on
a classical Markov process.
With these observations in mind, it is obvious, that the ENM master equation (or its two-qubits extension, see the "From one to two qubits dynamics and breaking also P-divisibility" and the 
"Bipartite GKSL representation" sections) needs not be related to any information backflow from dynamical environment.
The particular choice (\ref{gammas}) turns out to be a special case of this more general family of evolutions.

\section*{Markov dephasing dynamics}
%\label{MARKOV_DEPHASING_DYNAMICS}

To start with, consider simple dephasing dynamics of a qubit given by a master equation of GKSL
type \cite{luczka}
\begin{equation}\label{deph}
\dot{\rho}(t)=\sigma_{\alpha}\rho(t) \sigma_{\alpha} - \rho(t),
\end{equation}
where $\sigma_{\alpha} = \vec n_{\alpha} \cdot \vec \sigma$ is the Pauli matrix of some
direction $\vec n_{\alpha}$ ($|\vec{n}_{\alpha}|=1$). With
$\sigma_{\alpha}| \pm_{\alpha}\rangle = \pm|\pm_{\alpha}\rangle$,
Eq.~(\ref{deph}) leaves the populations
$\langle +_{\alpha}| \rho(t)|+_{\alpha} \rangle$ and $\langle -_{\alpha}| \rho(t)|-_{\alpha} \rangle$ constant,
the coherences $\langle +_{\alpha}| \rho(t)|-_{\alpha} \rangle$, $\langle -_{\alpha}| \rho(t)|+_{\alpha} \rangle$,
however, decay with a factor $\e^{-2t}$.

Since this CPT map is unital, the dynamics is of random unitary or {\it random external field} type
\cite{alicki,landau, helm,budini0,Kropf}. In fact,
a physical realisation of Eq.~(\ref{deph}) for pure initial states is easily obtained from a fluctuating field $\xi(t)$
driving the unitary Schr\"odinger dynamics:

\begin{equation}\label{ruschrodinger}
 i\partial_t |\psi(t) \rangle = \xi(t)\sigma_{\alpha}|\psi(t)\rangle.
\end{equation}
Indeed, if $\xi (t)$ represents Gaussian real white noise
with $ \eal \xi(t) \ear_{\xi} = 0$ and $\eal \xi(t)\xi(s) \ear_{\xi} = \delta(t-s)$, the noise-averaged state
$\rho(t)= \left\eal |\psi(t)\rangle \langle\psi(t)| \right\ear_{\xi}$ is a solution of (\ref{deph})
(see also Supplementary Information). With the unitary
$ U_{\xi}(t,0):=\e^{-i\int\limits_0^t \xi(s)ds\sigma_{\alpha}}$ we find for an arbitrary initial condition:
\begin{equation}
\rho(t) =  \eal U_{\xi}(t,0)\rho(0) U^+_{\xi}(t,0)\ear_{\xi}.
\end{equation}
As shown in Supplemetary Information, the noise average can easily be performed analytically to give the solution
of (\ref{deph}) in Kraus form
\begin{equation}\label{dephasing}
\rho(t) =  \frac{1}{2}\left( (1+e^{-2t})\rho(0) + (1-e^{-2t}) \sigma_{\alpha}\rho(0)\sigma_{\alpha}\right).
\end{equation}

\section*{Mixture of Markov dephasing dynamics}
%\label{sec:Dephasing}

\begin{figure*}[]
\includegraphics[width=1.0 \textwidth]{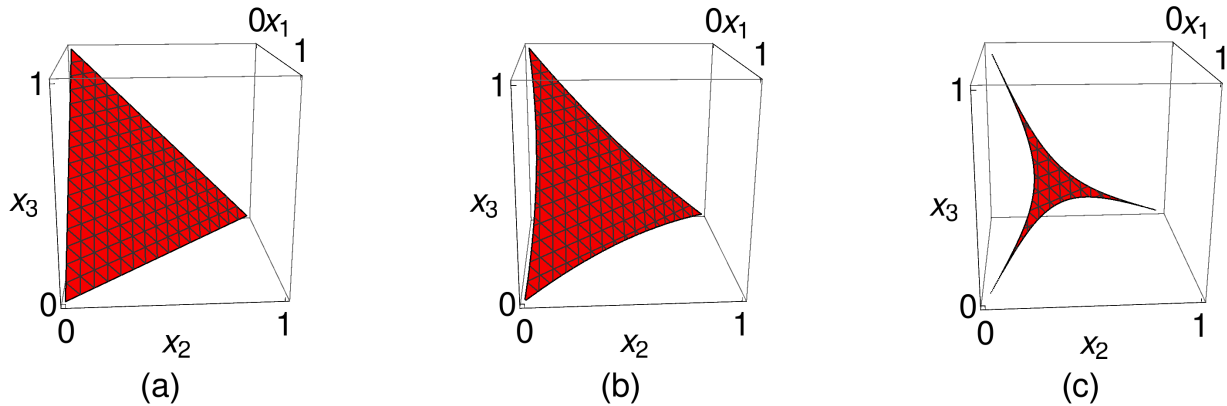}
\caption{Inner region: set of parameters $(x_1,x_2,x_3)$ with all three $\gamma_k(t)>0$ for some time t. (a) $t=0$ (defining triangle), (b) some later time $t>0$, (c)
$t\rightarrow \infty$ (asymptotic area).}
\label{bild-dreieck}
\end{figure*}

Now we allow the direction $\vec{n}_{\alpha}$ of the dephasing to be random with probability
distribution $p(\vec{n}_{\alpha})$. From (\ref{dephasing}) we see that with
$\vec n_\alpha = (n_1(\alpha),n_2(\alpha),n_3(\alpha))$, the averaged dynamics
depends on the second order correlations
\begin{equation}
  x_{kl}=\eal n_k(\alpha)n_l(\alpha)\ear_{\alpha}
\end{equation}
only.
Due to an overall orthogonal freedom of the whole problem, we may
assume a diagonal $(x_{kl})$ and will from now on use the notation
\begin{equation}
 x_k := \eal n_k^2(\alpha)\ear_{\alpha},
\end{equation}
assuming that $x_{kl}=0$ for $k\neq l$.
As final result, the dynamical map arising from averaging over noise $\xi(t)$ and
direction $\vec{n}_{\alpha}$ is again a map given in Kraus form by:
\begin{eqnarray}\label{rho-solution_diag}
\rho(t) &=& \eal U_{\xi}(t,0)\rho(0) U^+_{\xi}(t,0)\ear_{\xi,\alpha} \nonumber
 \\&=&\frac{1}{2} (1+e^{-2t})\rho(0) + \frac{1}{2}(1-e^{-2t})
 \sum\limits_{k=1}^3 x_k\sigma_k\rho(0)\sigma_k.\nonumber\\
\end{eqnarray}

The three positive parameters $x_1,x_2,x_3$, with $x_1+x_2+x_3 = 1$ (the Cartesian
variances of the distribution) are the only quantities of $p(\vec n_\alpha)$
that determine the dynamics. In Bloch representation this corresponds to a monotonic and (in general) anisotropic shrinking of the Bloch sphere (see Supplementary Information).

Therefore, it also follows that (\ref{rho-solution_diag}) can be obtained from a mixture
of just three orthogonal dephasing directions along the Cartesian axes. Accordingly, the
underlying dynamical map may be written as
a mixture of three Markov (semigroup) dynamical maps according to
\begin{align}\label{LLL}
\Lambda_t = x_1 e^{t \mathcal{L}_1} +  x_2 e^{t \mathcal{L}_2} + x_3 e^{t \mathcal{L}_3} \ ,
\end{align}
where $\rho(t)= \Lambda_t[\rho(0)]$, $ \mathcal{L}_k[\rho(t)] = \sigma_k \rho(t) \sigma_k - \rho(t)$ as in (\ref{deph}).
The variances may thus be seen as probabilities $x_k$ of choosing either of three
semigroup evolutions $e^{t \mathcal{L}_k}$ for the dynamics.

We conclude that {\it dephasing dynamics in random directions}
can be written in two ways as a mixture of CP-divisible maps. Representation (\ref{rho-solution_diag})
is a continuous mixture of unitary (Schr\"odinger) time evolutions, while in (\ref{LLL}) we have a discrete,
finite sum of irreversible Markov GKSL dynamics. As we will show next, the
corresponding master equation is just (\ref{ru-meq}), with possibly negative rates.\\

{\bf{Master equation and negativity of decay rates.}}
As shown in Supplementary Information, we find that the map $\Lambda_t$ from (\ref{LLL})
satisfies the time-local master equation
\begin{equation}
  \dot{\Lambda}_t = \mathcal{L}_t \Lambda_t,
\end{equation}
with the generator of the dynamics acting on density operators according to
\begin{equation}\label{}
  \mathcal{L}_t[\rho(t)] = \frac 12\sum_{k=1}^3 \gamma_k(t)(\sigma_k \rho(t) \sigma_k - \rho(t)) \ ,
\end{equation}
as in (\ref{ru-meq}).
The time dependent decoherence rates can be expressed as
\begin{eqnarray}
  \gamma_1(t)  &=& \left( \mu_1(t) - \mu_2(t) - \mu_3(t) \right), \nonumber \\
  \gamma_2(t) &=&  \left( -\mu_1(t) + \mu_2(t) - \mu_3(t) \right), \label{gammas3} \\
   \gamma_3(t) &=& \left( -\mu_1(t) - \mu_2(t) + \mu_3(t) \right), \nonumber
\end{eqnarray}
with
\begin{align}\label{}
  \mu_1(t) = - \frac{x_2+x_3}{x_2+x_3 + e^{2t} x_1}, &&
  \mu_2(t) = -  \frac{x_3 +x_1}{x_3+x_1 + e^{2t} x_2}, \nonumber \\
	\mu_3(t) = - \frac{x_1+x_2}{x_1+x_2 + e^{2t} x_3}\ . \nonumber
\end{align}
As we will work out in detail, these rates need not be positive. Thus,
the random mixture of Markovian dephasing leads to a time-local master
equation with possibly negative decay rates.\\

{\bf{Discussion of the negativity of the rates.}}
The parameter set of variances (or probabilities) $x_1,x_2,x_3$ with $x_1+x_2+x_3=1$ and $x_k$
positive represents a triangular area in 3-dimensional space spanned by the vectors
$\vec r = (x_1,x_2,x_3)$, see Fig.~\ref{bild-dreieck}. We refer to that set as
the parameter triangle. We display in Fig.~\ref{bild-dreieck} (hatched) that subset of parameters, for which
all $\gamma_k(t)$ are positive at that particular time t: (a) $t=0$, (b) some intermediate time $t>0$,
and (c) $t\rightarrow\infty$. Clearly, initially for $t=0$,
all $\gamma_k(0)=2x_k\ge 0 $ are non-negative. Later, only a symmetric triangular-star shaped
region near the centre reaching out to the tips of the parameter triangle corresponds to choices
of parameters for which all $\gamma_k(t)$ are
non-negative. Regions near the edges of the parameter triangle but away from the lines connecting the vertices with the centre of the triangle
correspond to choices of the $x_k$
that lead to a negative $\gamma_k(t)$ for some $t>t_*$. As $t\rightarrow \infty$,
an asymptotic finite area of that shape remains (we call it asymptotic area) for which all $\gamma_k(t)\ge 0$ for all times.
We will investigate the shape and size of that area in more detail later.

The rates have the following seven properties:
i) All rates start off non-negatively, $\gamma_k(0)=2x_k\ge 0$.
ii) At most one $\gamma_k(t)$ can turn negative.
iii) Once a $\gamma_k(t)$ turns negative at $t=t_*$, it remains negative ever after: $\gamma_k(t)<0$ for all $t>t_*$ (and the other two rates are always positive).
iv)  At the vertices of the defining triangle one of the $\gamma_k(t)$ equals $2$, the other two
equal $0$ and all three remain at those constant values (GKSL).
v)  All $x_1,x_2,x_3$ lying on the edges of the triangle (except vertices), i.e. when
exactly one of the $x_k=0$, give one of the $\gamma_k(t)<0$ {\it for all} $t>0$. The ENM master equation is
of that kind with $x_1=x_2=\frac{1}{2}$ and $x_3=0$. In those cases the dephasing is complete in that direction, and the corresponding probability distribution has a form 
$p(\vec{n}_{\alpha})=p(n_i,n_j)\delta(n_k).$
vi) For all parameters outside the asymptotic parameter area there exists some time $t_*>0$, so that for all $t<t_*$ all  $\gamma_k(t)$ are positive, and for all $t>t_*$
one
of the  $\gamma_k(t)$ is negative.
vii) We have $\gamma_1(t)+\gamma_2(t) \geq 0$ for all times (and cyclic) and thus,
the dynamics is P-divisible for all times and all choices of parameters \cite{darek,chruscinski,chen}.

We thus see that (quasi-)GKSL dynamics is only realised for our {\it dephasing in random directions}-process
for choices of $(x_1,x_2,x_3)$ within the asymptotic parameter area. Outside that area one of the
rates turns negative eventually (or immediately, for values at the border) and thus, the
corresponding CPT map is not CP-divisible. Remarkably, for all possible choices of parameters, the
map is P-divisible \cite{chruscinski}.

If the possible parameters $(x_1,x_2,x_3)$ are uniformly distributed over
the parameter triangle, the
probability for the corresponding {\it dephasing process in random directions} to be
of quasi-GKSL type is just the area of the asymptotic area relative to the
full parameter triangle.

As expanded in detail in Supplementary Information, in an appropriate
parametrization, the shape of the
asymptotic area is determined
by one of Newton's cubic curves \cite{Newton},
\begin{equation}
 x^2y+x-y=0.
\end{equation}
For the relative area of parameters outside the asymptotic (hatched) area, we find
\begin{align}
 \frac{A_{\rm {\text{non-CP-div}}}}{A_{\rm{\text{tot}}}} =
 \int\limits_0^{\sqrt{5}-2}dx \frac{6(3-3x-3x^2-x^3)x}{\sqrt{1-\frac{4x}{1-x^2}}(1-x^2)(1+x)}
 \approx 0.87,
\end{align}
see Supplementary Information.
Interestingly, only $13\% $ of all {\it dephasing in random directions} dynamical maps are CP-divisible
or of quasi-GKSL type. In particular, near the tips of the triangles, as the sides turn into tangents,
only a vanishingly small set of CP-divisible maps remains for small fluctuations around a Cartesian direction.
Thus, dephasing in one of the Cartesian directions with only the slightest fluctuations around that direction
leads to a dynamics with negative dephasing rate with an overwhelming probability.

\section*{Memory kernel master equation}
\label{kernel}
It is worth noting that the dynamics (\ref{rho-solution_diag}) can also be described with a
master equation involving a memory kernel \cite{nakajima,zwanzig}
\begin{equation}\label{kernel}
\dot{\rho}(t)=\int\limits_0^t K(t-s)\rho(s)ds.
\end{equation}
For our dynamics, we find a kernel of the following form:
\begin{equation}\label{}
  K(t-s)\rho(s) = \frac 12 \sum_{k=1}^3 K_k(t-s)( \sigma_k \rho(s) \sigma_k - \rho(s)),
\end{equation}
with
\begin{equation}
  K_k(t)  = x_k \delta(t) + \eta_k(t)
%, &&
%  K_2(t) =  x_2 \delta(t)  + \eta_2(t), \nonumber \\   &&
%  K_3(t) = x_3 \delta(t) + \eta_3(t)  \ ,
\end{equation}

and
\begin{eqnarray}
  \eta_1(t)  &= &\frac 12  \left( X_1(t)- X_2(t) - X_3(t) \right), \nonumber \\
  \eta_2(t)  &= &\frac 12  \left( -X_1(t) + X_2(t) - X_3(t) \right), \nonumber \\
  \eta_3(t)  &= &\frac 12  \left( -X_1(t) - X_2(t) + X_3(t) \right)  \ ,
\end{eqnarray}
with $X_k(t) =x_k(1-x_k) e^{-x_kt}$.
Hence
\begin{align*}
K(t)=&\frac{1}{2}\left(x_1 \mathcal{L}_1 + x_2 \mathcal{L}_2 + x_3 \mathcal{L}_3 \right)\delta(t)
	\\  &+ \frac{1}{2} \left( \eta_1(t)\mathcal{L}_1 + \eta_2(t)\mathcal{L}_2 + \eta_3(t) \mathcal{L}_3 \right) .
\end{align*}
Interestingly, the memory kernel $K(t)$ has the following structure
\begin{equation}\label{}
  K(t) = K_{\rm loc} \delta(t) + K_{\rm nloc}(t) ,
\end{equation}
where the time-local part $K_{\rm loc}= \frac 12 (x_1 \mathcal{L}_1 + x_2 \mathcal{L}_2 + x_3 \mathcal{L}_3)$  is just the weighted sum of the three Cartesian GKSL
dephasing generators. The non-local part depends on three smooth functions $\eta_k(t)$.

As observed in ref. \citenumns{chruscinski3} and confirmed here, a local in time master equation description
of the dynamics has complementary properties to a memory kernel master equation, in the sense that a "nice"
functional form in one formulation may lead to a more singular description in the other.

We see that the mixture of Markovian dephasing dynamics studied in this paper `` $ x_1 e^{t \mathcal{L}_1} +  x_2 e^{t \mathcal{L}_2} + x_3 e^{t \mathcal{L}_3} $'' may well
be written in a form involving a ``memory integral'', that is, apart from the more or less clear local term $K_{\rm loc}$ it contains a truly non-local part $K_{\rm nloc}(t)$.
In open quantum system dynamics, non-local master equations of type \eqref{kernel} appear naturally from a dynamical environment, as, for instance, in the Nakajima-Zwanzig approach 
\cite{nakajima,zwanzig}. Obviously, no dynamical environment exists in our constructions.

\section*{Classical Markov process representation of dynamics}
%\label{sec:classicmarkovchain}
So far we have acknowledged that the simple mixture of Markovian dynamics may well
lead to a master equation involving negative rates. Remarkably, as we will explain
in this section, that latter master equation may easily be simulated using a classical Markov process.

We start with the Kraus representation of the {\it dephasing dynamics in
random directions}, Eq.~(\ref{rho-solution_diag}).
We introduce the unitarily transformed states
$\rho_k:=\sigma_k \rho(0) \sigma_k$, $k=0,...,3$ (with $\sigma_0=\opone$)
and corresponding probabilities $p_k(t)$ such that the state at time $t$
reads $\rho(t) = \sum\limits_{k=0}^3 p_k(t) \rho_k$. The probabilities
\begin{align}
p_0(t)=\frac{1}{2}\left(1+\e^{-2t} \right), && p_k(t)=\frac{x_k}{2}\left(1-\e^{-2t} \right) \label{solp}
\end{align}
can be read off from the Kraus representation (\ref{rho-solution_diag}).

\begin{figure}[]
\includegraphics[width=0.4 \textwidth]{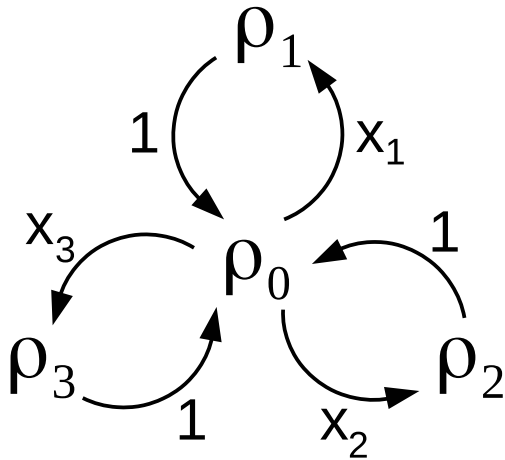}
\caption{Graphical representation of a master equation \eqref{pcapital} with $\rho_j=\sigma_j \rho \sigma_j$, with $\sigma_0=\opone$, where jumps
$\rho_0 \rightarrow \rho_1$ occur with rate $\Gamma_{01}$ ($=x_1$),
  and $\rho_1 \rightarrow \rho_0$ with rate  $\Gamma_{10}$ ($=1$) , etc. No jumps $\rho_1 \rightarrow \rho_2$, $\rho_2 \rightarrow \rho_1$ nor
  $\rho_1 \rightarrow \rho_3$,...,
   take place.  }
\label{bild1}
\end{figure}

As elaborated upon in Supplementary Information, these probabilities are solutions of
the rate equations
\begin{align}\label{pcapital}
  \frac{d}{dt} \left( \begin{array}{c} p_0(t) \\ p_1(t)  \\ p_2(t)  \\ p_3(t)  \end{array} \right) =  \left( \begin{array}{cccc} - 1  & 1
    & 1  & 1  \\   x_1  & - 1 & 0  & 0  \\   x_2  & 0  & - 1  & 0  \\   x_3  & 0
     & 0  & - 1  \end{array} \right) \left( \begin{array}{c} p_0(t)  \\ p_1(t)  \\ p_2(t)  \\ p_3(t)  \end{array} \right)
\end{align}
that are of the form of a classical Pauli master equation \cite{vanKampen}
\begin{equation}\label{classical_master}
 \dot{p}_k(t) = \sum_j \Big( \Gamma_{j\rightarrow k}p_j(t) - \Gamma_{k\rightarrow j}p_k(t) \Big) ,
\end{equation}
with {\it positive} an {\it time-independent} rates $\Gamma_{0 \rightarrow k }=x_k, \Gamma_{k\rightarrow 0}=1$,
and all other rates being zero.
The corresponding transitions are displayed in Fig.~\ref{bild1}.

Most remarkably,
despite the negativity of the rates of the underlying quantum master equation, its
solution $\rho(t)$ can be obtained from the classical Markov master
equation (\ref{classical_master})
 according to the following construction.
Take a classical process between four classical states $\{r_0,r_1,r_2,r_3\}$ as determined
from the classical master equation (\ref{classical_master}). For a transition from
state  $r_0$ to some $r_k$ (with $k=1,2,3$), apply the unitary transformation
$\sigma_k$ to the state, so that $\rho_0\rightarrow \rho_k$ occurs with rate
$\Gamma_{0\rightarrow k}=x_k$.
Equally, if a jump from $r_k$ ($k=1,2,3$) back to $r_0$ occurs, again apply the unitary $\sigma_k$ to
the current state so that $\rho_k \rightarrow \rho_0$ with rate  $\Gamma_{k\rightarrow 0}=1$.
No other jumps can take place, see Fig.~\ref{bild1}.

By construction, $\rho(t)=\sum\limits_{k=0}^3p_k(t)\rho_k$ is the solution (\ref{rho-solution_diag}).
Consequently, one can also simply generate the probability distribution $p_k(t)$ simulating the classical
Markov process and afterwards accordingly mix the final density matrix using the four $\rho_k$.

We have managed to describe the process \eqref{ru-meq} based on the classical master equation \eqref{classical_master}
with positive, time independent rates. So we find a Markov chain representation of ENM.\\

{\bf{Negative rate classical master equation.}}
Starting from the time-local master equation (\ref{ru-meq}) and writing its solution in the form of the dynamical map
\begin{equation}\label{dynmap}
  \Lambda_t[\rho(0)] = \sum_{j=0}^3 P_j(t) \sigma_j\rho(0) \sigma_j ,
\end{equation}
we obtain the following equation for the probability 4-vector $\vec{P}(t)$ (for clarity we suppress the time dependence of $\gamma_j(t)$):
\begin{align}\label{11}
  \frac{d}{dt} \left( \begin{array}{c} P_0(t) \\ P_1(t)  \\ P_2(t)  \\ P_3(t)  \end{array} \right) &= \frac 12 \left( \begin{array}{cccc} - \gamma_0  & \gamma_1
    & \gamma_2  & \gamma_3  \\   \gamma_1  & -\gamma_0  & \gamma_3  & \gamma_2  \\   \gamma_2  & \gamma_3  & -\gamma_0  & \gamma_1  \\   \gamma_3  & \gamma_2
     & \gamma_1  & -\gamma_0  \end{array} \right) \left( \begin{array}{c} P_0(t)  \\ P_1(t)  \\ P_2(t)  \\ P_3(t)  \end{array} \right) ,
\end{align}
where $\gamma_0(t)  := \gamma_1(t)  + \gamma_2(t)  + \gamma_3(t) $. It can be rewritten in the form of a Pauli master equation
\begin{align}\label{jump-neg}
\dot{P}_k(t)=\frac{1}{2}\sum\limits_j \left(\gamma_{j\rightarrow k}(t)P_j(t)-\gamma_{k\rightarrow j}(t)P_k(t) \right),
\end{align}
with $\gamma_{0\rightarrow j}(t)=\gamma_{j\rightarrow 0}(t)=\gamma_j(t)$, $\gamma_{k\rightarrow j}(t)=\gamma_l(t)$, for $k\neq j \neq l$ ($k,j,l=1,2,3$).
As for the quantum master equation the transition rates can turn negative, equation \eqref{jump-neg} does not define a proper Markov process.

The solution
\begin{equation*}
  \vec{P}(t)=T(t)\vec{P}(0)
\end{equation*}
can be obtained from the propagator $T(t)$ as given in Supplementary Information. For the initial
 condition $\vec{P}(0)=(1,0,0,0)^T$ we find
 positive $P_k(t)$ for all $t$. Thus, despite the negative rates, the master equation \eqref{jump-neg} defines a proper evolution for a probability distribution for
 that particular choice of $\vec{P}(0)$.
 Similarly, for that initial condition only, we have $\Lambda_0=\opone$.\\
 Due to the negative rates one is tempted to think of \eqref{jump-neg} as representing a non-Markovian jump process. Yet, it is clear that $\vec{P}(t)=\vec{p}(t)$ for 
 initial condition
 $(1,0,0,0)^T$, and $\vec{P}(t)$ is therefore also a solution of a Markovian jump process \eqref{classical_master}. Hence, one should also be careful with the 
 interpretation
 of classical master equations involving negative rates.\\

{\bf{Special case.}}
For the special choice of $\gamma_k(t)$ given in Eq.~(\ref{gammas}) (introduced in ref. \citenumns{hallcresserandersson}) and assuming $P_0(0)=1$ and $P_k(0) = 0$ one finds
\begin{eqnarray}
  P_0(t) &=& \frac 12 (1+e^{-2t}) \\
  P_1(t) &=& P_2(t) = \frac 14 (1-e^{-2t})  \\
  P_3(t) &=& 0 \ .
\end{eqnarray}
Hence $P_3(t)$ is irrelevant and the solution is generated from (\ref{pcapital}) via the simplified Markov chain:

\begin{equation}\label{J}
  \frac{d}{dt} \left( \begin{array}{c} P_0(t) \\ P_1(t) \\ P_2(t)  \end{array} \right) =  \left( \begin{array}{ccc} - 2 & 2 &
  2 \\   1 & -2 & 0 \\  1 & 0 & - 2 \end{array} \right) \left( \begin{array}{c} P_0(t) \\ P_1(t) \\ P_2(t)  \end{array}
  \right),
\end{equation}
with positive $\gamma_1=\gamma_2=1$.
It is evident that (\ref{J}) generates a Markov semigroup.
For a discussion for different initial conditions see Supplementary Information.\\

{\bf{Realisation with orthogonal states.}}
%\label{orth-NCP-div}
Note that the $\rho_j=\sigma_j \rho(0) \sigma_j$ are not mutually orthogonal, so they cannot be
distinguished faithfully by a measurement. However, a truly classical
implementation involving four classical (i.e. orthogonal) states can be found by
expanding the dimension of our system to four qubits
($H=H^A\otimes H^B=\mathbb{C}^2\otimes \mathbb{C}^8$).

For this construction, we define the following extended dynamics involving the three ancilla
qubits:
\begin{align}\label{church}
\dot{\tilde{\rho}}(t)=\frac{1}{2}\sum\limits_{k=1}^{3} \gamma_k(t)((\sigma_k\otimes U_k)\tilde{\rho}(t)(\sigma_k\otimes U^+_k) - \tilde{\rho}(t)),
\end{align}
where $U_k$ are unitary operators, specified below.\\
Tracing out the ancilla (B) degrees of freedom, this dynamics reduces to (\ref{ru-meq}).

To construct the four orthonormal states we write the initial density operator in diagonal form:
\begin{align*}
\rho_0= p_1 |\phi_1\rangle \langle \phi_1 |+p_2 |\phi_2\rangle \langle \phi_2 |,
\end{align*}
with orthonormal vectors $|\phi_1\rangle,|\phi_2\rangle$ and non-negative
probabilities $p_1,p_2=1-p_1$.
Our four-qubit states are defined in the following way:
\begin{align*}
| \Psi_0 \rangle&= \sqrt{p_1}|\phi_1\rangle|\psi_1\rangle +\sqrt{p_2}|\phi_2\rangle|\psi_2\rangle,\\
| \Psi_k \rangle&= \sqrt{p_1}\sigma_k\otimes  U_k|\phi_1\rangle |\psi_{1}\rangle +\sqrt{p_2}\sigma_k\otimes U_k|\phi_2\rangle |\psi_{2}\rangle,
\end{align*}
where the $U_k$ are chosen, such that $U_k|\psi_i\rangle=|\psi_{2k+i}\rangle$ and $|\psi_{l}\rangle \in  \mathbb{C}^8$, $l=1,...,8$ are mutually orthogonal
and normalized.

These four vectors of course don't build a basis of the $\mathbb{C}^{16}$. Nonetheless, if we
set $| \Psi_0 \rangle \langle  \Psi_0 |$ as the initial state of our four-qubit
system and let it evolve according to \eqref{church}, the output state is always a mixture of these four
states.

Consequently, we get a realisation of the dynamics (\ref{ru-meq}) with distinguishable states.
In a lab, therefore, one might choose to measure in a time-continuous fashion the actual
four-qubit state such as to have a time-continuous (Markov) realisation of the classical
process described in Fig.~\ref{bild1}. By construction, the ensemble mean of the corresponding reduced states,
at all times of continuous monitoring,
is a solution of the original negative-rate master equation.\\

{\bf{From one to two qubits dynamics and breaking also P-divisibility.}}
%\label{orth-NP-div}
From the non-CP-divisibility of the one qubit dynamical map \eqref{LLL} one can conclude that the corresponding map for two qubits, where the first qubit undergoes the
dynamics \eqref{LLL} and the second one is frozen, is not P-divisible. Nonetheless, also in this case we can find a classic Markov process representation, which can be
realised with orthogonal states. \\
To show this we expand the initial state of the two-qubit system in a following form:
\begin{align}\label{initial}
\rho_{AB}(0)=\sum\limits_{ikmn=1}^2a_{ikmn}|\varphi_i\rangle \langle \varphi_k| \otimes |\psi_m\rangle \langle \psi_n|,
\end{align}
where $|\varphi_1\rangle$, $|\varphi_2\rangle$ are the eigenstates of the first qubit $A$ (with the corresponding eigenvalues $p_1$, $p_2$, $p
_1+p_2=1$) and $|\psi_1\rangle$, $|\psi_2\rangle$ are two orthogonal states of the second qubit $B$. Equation \eqref{initial} represents a general initial state, also
entangled states are included. \\
The coefficients $a_{ikmn}$ are some complex numbers, which have to satisfy
\begin{align*}
\rho_A(0)=Tr_B(\rho_{AB}(0))= p_1 |\varphi_1\rangle \langle \varphi_1|+p_2 |\varphi_2\rangle \langle \varphi_2| && \Leftrightarrow
\end{align*}
\vspace{-0.5cm}
\begin{align}
 \sum\limits_{l=1}^2 a_{11ll}=p_1, && \sum\limits_{l=1}^2 a_{22ll}=p_2, &&
\sum\limits_{l=1}^2 a_{12ll}=\sum\limits_{l=1}^2 a_{21ll}=0, \label{1twoqb}
\end{align}
\vspace{-0.5cm}
\begin{align}
\rho_{AB}(0)=\rho^+_{AB}(0) && \Leftrightarrow && a_{ikmn}=a^*_{kinm} \label{3twoqb}.
\end{align}
For the initial jump state in the extended Hilbert space (by a third system $C$) we make an ansatz:
\begin{align*}
|\xi_0\rangle=\sum\limits_{ik=1}^2\sum\limits_{l=1}^4 c_{ikl}|\varphi_i\rangle|\psi_k\rangle|\chi_l\rangle,
\end{align*}
where $|\chi_l\rangle$ are mutually orthogonal. The 16 coefficients $a_{ikmn}$ are mapped on the 16 coefficients $c_{ikl}$ with
$a_{ikmn}=\sum\limits_{l=1}^4 c_{iml}c^*_{knl}$,
following from $\rho_{AB}(0)=Tr_C(\rho_0)=Tr_C(|\xi_0\rangle\langle \xi_0|)$.\\
Per construction, Eq. \eqref{3twoqb} is fulfilled, also the positivity of the $p_1$, $p_2$ is guaranteed for all $c_{ikl}$. The other conditions for $a_{ikmn}$
put some constraints on the possible choice of $c_{ikl}$.\\
The other jump states take the form ($j=1,2,3$):
\begin{align*}
|\xi_j\rangle=\sum\limits_{ik=1}^2\sum\limits_{l=1}^{4} c_{ikl} (\sigma_j\otimes\opone\otimes V_j)|\varphi_i\rangle|\psi_k\rangle|\chi_l\rangle,
\end{align*}
where the unitary $V_j$ are chosen, such that $V_j|\chi_l\rangle=|\chi_{4j+l}\rangle$ and $|\chi_{i}\rangle \in  \mathbb{C}^{16}$, $i=1,...,16$ are mutually orthogonal
and normalized. Consequently, $|\xi_0\rangle,...,|\xi_3\rangle$ are mutually orthogonal. To achieve this we have to extend our Hilbert space by four qubits, so overall our
system consists of six qubits.\\
Fulfilment of condition \eqref{1twoqb} guarantees that $Tr_{B,C}(\rho_j)=Tr_{B,C}(|\xi_j\rangle\langle\xi_j|)=\sigma_j \rho_A(0)\sigma_j$.
In addition, the state of the second B qubit is the same
for all $|\xi_0\rangle,...,|\xi_3\rangle$.\\
Accordingly, also the dynamics of two qubits, where the first undergoes \eqref{LLL} and the second is frozen, can be mapped on the (time-continuous limit of the)
Markov jump process graphically represented in Figure \ref{bild1}, where the states $\rho_k$ are redefined.
From this we conclude, that there are non-P-divisible maps, for which a classical Markov process description is possible. Therefore, both non-CP-divisibility
\cite{chruscinski}, but also the weaker non-P-divisibility \cite{breuer5}, are questionable indicators for the occurrence of memory effects associated with dynamics of 
environmental degrees of freedom.\\

{\bf{Bipartite GKSL representation.}}
%\label{bipart}
Interestingly,  the dynamics defined by \eqref{rho-solution_diag} may be represented via
\begin{align}\label{bipartite}
\Lambda_t[\rho(0)]=\Tr_E(\e^{t\mathcal{L}}[\rho(0)\otimes \rho_E]) ,
\end{align}
where $\mathcal{L}$ denotes a time independent bipartite GKSL generator. This construction is based on the {\em correlated projection method} \cite{breuer6,budini1}:
one defines the initial state of the bipartite system to be the following quantum-classical state
\begin{equation}\label{QC}
  \tilde{\rho}(0) = \sum_{i=1}^3 \rho_i(0) \otimes  |i\>\<i| ,
\end{equation}
where $|i\>$ are orthonormal vectors in $\mathcal{H}_E = \mathbb{C}^3$. Suppose now that the generator $\mathcal{L}$ gives rise to  $e^{t \mathcal{L} }\tilde{\rho}(0) = \sum_{i=1}^3 \rho_i(t) \otimes  |i\>\<i|$, that is, the bipartite evolution preserves the structure (\ref{QC}). Then  the partial trace $\rho(t) = \Tr_E \tilde{\rho}(t)$ is defined by $\rho(t) = \sum_i \rho_i(t)$. Note that in general this prescription does not define a dynamical map \cite{breuer6}. However, if $\rho_i(0) = x_i \rho(0)$, then   (\ref{QC}) defines  a product state $\rho(0)\otimes \rho_E$, with $\rho_E = \sum_i x_i |i\>\<i|$  and hence one arrives at the legitimate map (\ref{bipartite}). 

Let us define $\mathcal{L}$ by
\begin{equation}\label{total}
  \mathcal{L}[\tilde{\rho}] = \sum\limits_{k=1}^3\left[ {\tilde{\sigma}_k}\tilde{\rho} {\tilde{\sigma}_k}- {\tilde{\rho}}\right],
\end{equation}
where $\tilde{\sigma}_i=\sigma_i\otimes P_i$ and $P_i = |i\>\<i|$. One immediately finds 
\begin{equation}\label{}
  \rho(t) =\Lambda_t[\rho(0)] = \sum_{k=1}^3 x_k e^{t \mathcal{L}_k} \rho(0) . 
\end{equation}
 Such a bipartite Markovian dynamics, which potentially gives rise to the non-Markovian evolution on the reduced level, was already widely described
in the recent literature, e.g. in ref. \citenumns{budini1,Ciccarello,budini2,budini3,Kretschmer}.
Notice however the qualitative difference of our description to the cited one: as is apparent from \eqref{total} in our case the dynamics of the ancilla state is frozen
(the reduced density matrix of the ancilla does not change) and  there is never any entanglement
between the system and an ancilla. That means that the ancilla is only a "casual bystander" during the whole dynamics $t>0$. Consequently, it is hard to see any
information backflow in this construction.

The corresponding GKSL master equation also exists in the extended two qubits case:
\begin{align}
\dot{\hat{\rho}}(t)=\sum\limits_{k=1}^3\left[ \hat{\sigma}_k\hat{\rho}(t) \hat{\sigma}_k
-\hat{\rho}(t)\right],
\end{align}
where $\hat{\sigma}_k=\sigma_k\otimes\opone\otimes P_k$ and $\hat{\rho}(t)=\sum\limits_{k=1}^3 x_k\tilde{\rho}_k(t)\otimes P_k$, with
$\tilde{\rho}_k(t)=\left(\e^{t\mathcal{L}_k}\otimes \opone \right)[\tilde{\rho}(0)]$.
The dynamics of the first qubit is defined by \eqref{rho-solution_diag}, the second one and the ancilla state are frozen.
Notice, that the initial state of the two qubits can be chosen arbitrarily.\\
Also here the ancilla is only a "casual bystander" during the whole evolution $t>0$.

Actually, as can be easily seen from the above construction, such an embedding in a bipartite GKSL equation with a "casual bystander" ancilla is possible for all dynamics,
which can be written as a time-independent mixture of GKSL evolutions.

\section*{Conclusions}

This paper analyses a class of qubit evolutions $\Lambda_t[\rho] = \sum_{k=0}^3 p_k(t) \sigma_k \rho \sigma_k$ which can be written as a convex combination of Markovian
semigroups $\Lambda_t = x_1 e^{t \mathcal{L}_1} + x_2 e^{t \mathcal{L}_2} +x_3 e^{t \mathcal{L}_3}$, where $\mathcal{L}_k$ is a purely dephasing generator defined by
$\mathcal{L}_k[\rho] = \sigma_k \rho \sigma_k - \rho$. $\Lambda_t$ satisfies a time-local master equation, whose corresponding generator
$\mathcal{L}_t[\rho] = \sum_{k=1}^3 \gamma_k(t) (\sigma_k \rho \sigma_k-\rho)$ may contain exactly one decoherence rate $\gamma_k(t)$ which is negative for $t>t_*$.
Based on the concept of CP-divisibility such evolution is immediately classified as non-Markovian. Interestingly, within this class the evolution is P-divisible and hence
Markovian according to the concept of information flow \cite{breuer}. This is, therefore, another example showing that these two concepts do not coincide. Equivalently,
$\Lambda_t$ satisfies memory kernel master equation with the memory kernel $K(t)$ possessing apart  from the local part $(x_1 \mathcal{L}_1 + x_2  \mathcal{L}_2 + x_3
\mathcal{L}_3)\delta(t)$ a non-trivial non-local term suggesting the presence of memory effects.

More interestingly, however, we showed that $\Lambda_t$ may be easily realised  as stochastic averaging of the purely unitary evolution governed by dephasing dynamics in
random directions. Alternatively, there is a realisation based on a classical Markov process, where the probabilities $p_k(t)$ are governed by a classical Pauli master
equation. Such a classical Markov representation exists also for a non-P-divisible dynamics of an extended two qubit system.
In both cases a description with a bipartite GKSL equation, where the ancilla state is frozen, is possible, too.
These realisations show that actually there is no room for physical memory effects. 
This proves that the interpretation of both time-local and memory kernel master equations with respect to memory effects is a delicate issue.
A reduced description may not suffice to study the physics of memory in terms of information flow.\\

\vspace{1cm}

{\bf Acknowledgements} 
The authors would like to thank Kimmo Luoma and Anna Costa for fruitful discussions.
DC was partially supported by the National Science Centre project 2015/17/B/ST2/02026.
JP acknowledges funding from Academy of Finland (project 287750) and Magnus Ehrnrooth Foundation.\\

{\bf Author contributions statement} 
All the authors have contributed equally to developing the main ideas and discussing the results. The manuscript was written by N.M. and W.S. with
input from D.C. and J.P. N.M prepared the figures. The project was supervised by W.S.\\

{\bf Additional Information}
The authors declare that they have no competing financial interests.

\end{document}